\documentclass{segabs}

\usepackage[utf8]{inputenc}
\usepackage{amsmath}
\usepackage{amssymb}
\usepackage{bm}
\usepackage{longtable}
\usepackage{subfigure}
\usepackage{float}
\usepackage{caption}
\usepackage{amsmath,amsthm,amssymb,amsfonts}

\usepackage{mathptmx}
\DeclareMathAlphabet{\mathcal}{OMS}{cmsy}{m}{n}
\DeclareSymbolFont{largesymbols}{OMX}{cmex}{m}{n}

\renewcommand{\tensor}[1]{\mathcal{#1}}
\newcommand{\Rset}[4]{\mathbb{#1}^{#2\times #3\times #4}}
\newcommand{\XinRset}[5]{\tensor{#1} \in \mathbb{#2}^{#3\times #4\times #5}}
\newcommand{\POmega}[1]{\mathcal{P}_\Omega(#1)}
\newcommand{\XOmegaTilde}[1]{\widetilde{\tensor{#1}}_\Omega}
\newcommand{\XTilde}[1]{\widetilde{\tensor{#1}}}

\newcommand{\XOmegaLateralTilde}[2]{\XOmegaTilde{#1}^{(#2)}}
\renewcommand{\vector}[1]{\mathbf{#1}}
\renewcommand{\matrix}[1]{\mathit{#1}}
\newcommand{\argmin}[4]{\mathop{\mathrm{arg min}}\limits_{#1\in \mathbb{R}^{#2 \times #3 \times #4}}}
\newcommand{\argmina}[3]{\mathop{\mathrm{arg min}}\limits_{#1\in \mathbb{R}^{#2 \times #3}}}

\begin{document}

\title{Exact 3D seismic data reconstruction using Tubal-Alt-Min algorithm }

\renewcommand{\thefootnote}{\fnsymbol{footnote}} 

\author{Feng Qian$^{1}$, Quan Chen$^{2}$, Ming-Jun Su$^{3}$, Guang-Min Hu$^{2}$, Xiao-Yang Liu\footnotemark[1] \\
		$^{1}$Center of Information Geoscience, University of Electronic Science and Technology of China, \\
		$^{2}$School of Rresources and Environment, University of Electronic Science and Technology of China, \\
	    $^{3}$PetroChina Research Institute of Petroleum Exploration and Development (RIPED)-Northwest,  Lanzhou, China,\\
		\footnotemark[1]Department of Electrical Engineering, Columbia University, NY, USA
	}

%\footer{Example}
%\lefthead{Dellinger \& Fomel}
%\righthead{SEG abstract example}

\maketitle

\begin{abstract}
Data missing is an common issue in seismic data, and many methods have been proposed to solve it. 
In this paper, we present the low-tubal-rank tensor model and a novel tensor completion algorithm to recover 3D seismic data. 
This is a fast iterative algorithm, called Tubal-Alt-Min which completes our 3D seismic data by exploiting the 
low-tubal-rank property expressed as the product of two much smaller
tensors. Tubal-Alt-Min alternates between estimating those two tensor using least squares minimization. 
We evaluate its reconstruction performance both on synthetic seismic data and land data survey.
The experimental results show that compared with the tensor nuclear norm minimization algorithm, Tubal-Alt-Min 
improves the reconstruction error by orders of magnitude.
\end{abstract}

\section{Introduction}
Seismic data quality is vital to various geophysical processing. However, due to the financial and physical constraints, 
the real seismic survey data are usually incomplete. Seismic data reconstruction is a complex problem, and a large 
number of researchers have devoted themselves to the research of this field. Consequently, many approaches have 
been proposed to handle this problem. From the point of view of data organization, these methods can be divided into two categories.

The low dimensional based methods, such as transform-based methods which utilize the properties of the seismic trace in an auxiliary 
domain \cite[]{hennenfent2010nonequispaced}.  Liner prediction theory  use the predictability of the signal in the f‐x or t‐x 
domain \cite[]{naghizadeh2007multistep}. And methods which exploit the low-rank nature of seismic data embedded in 
Hankel matrices \cite[]{Oropeza2011Simultaneous}. 

Those low dimensional based methods usually ignores the spatial structure of the seismic traces. 
While the spatial structure coherence is very important for the seismic data completion, hence recent developments in high dimensional tensor 
completion approaches exploit various tensor decomposition model are widely used in seismic data reconstruction \cite[]{Kreimer2013Restricted, Gregory20155D}. 
Those high-order tensor decomposition approaches have became a trend for seismic data completion, 
and the exist different definitions for tensor decomposition that lead to different tensor completion
model, i.e. the higher-order singular value decomposition (HOSVD) \cite[]{Kreimer2011A}, the 
tuker decomposition \cite[]{Silva2012Hierarchical}, and tensor SVD (tSVD) decomposition
\cite[]{Ely20155D}.

In this paper, we focus on a new seismic data reconstruction algorithm which based on low-tubal-rank tensor decomposition model
possesses extremely high precision seismic data recovery performance. Because of the high redundancy or coherence between one seismic trace
to the others \cite[]{Gregory20155D}, we assume that the fulled sampled seismic volume has low-tubal-rank property in the tSVD domain. Therefore, 
we can solve the seismic tensor completion problem through an alternating minimization algorithm for low-tubal-rank tensor completion (Tubal-Alt-Min) \cite[]{Liu2016Low-tubal-rank, liu2016low}. We have evaluated the performance of this approach on both synthetic and field seismic data.

\section*{Notations}

The data in seismic survey is a natural high-dimensional tensor, such as the 3D poststack seismic data which consists of 
one time or frequency dimension and two spatial dimensions corresponding to xline and inline directions. 

Throughout the paper, we denote those 3D seismic tensor in time domain by uppercase calligraphic letter, $\XinRset{\tensor{T}}{R}{m}{n}{k}$,
and denote the frequency domain 3D seismic tensor by $\XTilde{T} \in \Rset{R}{m}{n}{k}$ correspondingly.
Uppercase letter $\matrix{A} \in \mathbb{R}^{m \times n}$ denotes matrix, and lowercase boldface 
letter $\vector{x} \in \mathbb{R}^n$ denotes vector. Let $[n]$ denotes the set $\{1,2,\dots,n\}$. 
In addition, we introduce an important tensor operator $\textit{t-product}$ \cite[]{MISHA2013THIRD}.

$\textit{t-product}.$ The tensor-product $\tensor{T} = \tensor{X} \ast \tensor{Y}$ of $\tensor{X} \in \Rset{R}{n_1}{n_2}{k}$ and
$\tensor{Y} \in \Rset{R}{n_2}{n_3}{k}$ is a tensor of size ${n_1} \times {n_2}\times k$, 
$\tensor{T} (i, j, :) = \sum_{s=1}^{n_2}\tensor{X}(i,s,: ) \ast \tensor{Y}(s, j, :)$, for $i \in [n_1]$ and $j\in [n_3]$.

\section{Problem Setup}
From what we have stated above, we know that seismic data comprises many traces that provide a spatio-temporal 
sampling of the reflected wavefield. However, caused by various factors, such lost many important informations.
Such as the exist of reservoir, residential or any other obstacle in the seismic data acquisition areas will lead to 
under-sampled seismic record. The missing traces will complicate certain data processing steps such as the prediction 
accuracy of underground reservoirs. Hence, the completion step in seismic data processing is of grate significance. 

In this paper, we explored the relationship between low-tubal-rank and under-sampled rate firstly, and found that 
low-tubal property is positively correlated with the sampling rate. Figure 1 shows the detail experiment result.
Base on this work, we as-
\begin{figure}[H]
	\centering
	\vspace{-0.4cm}
	\includegraphics[width=8cm]{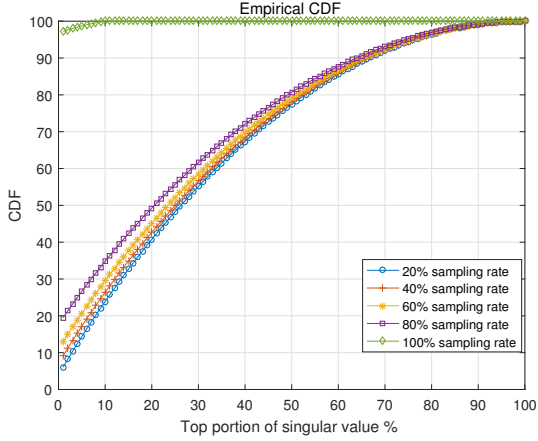}
	\caption{The empirical CDF of singular value. }
\end{figure}

\rule{7.9cm}{0.12em} \\
\textbf{Alg. 1} Tubal Alternating Minimization: AM($\POmega{\tensor{T}}$, $\Omega$, $L$, $r$)
\rule{7.9cm}{0.08em}
\textbf{Input:} Observation set  $\Omega \in [n] \times [m] \times [k] $ and the corresponding elements $ \POmega{\tensor{T}}$ 
, number of iterations $L$, target tubal-rank $r$. 

1: $\ \ \tensor{X}_0 $ $\gets$ Initialize($\POmega{\tensor{T}}, \Omega, r$),   \\
2: $\ \ \XTilde{X}_0 $ $\gets$ fft($ \tensor{X}_0 $, [], 3), \ \ $\XOmegaTilde{T} \gets$ fft($ \tensor{T}_\Omega$, [], 3), \\
$~~~~~~~~\XOmegaTilde{P}$ $\gets$ fft($\tensor{P}_\Omega$, [], 3), \\
3:	\textbf{\ \ \ for} $l$ = 1 to $L$ \textbf{do} \\
$~~~~~~~~~~~~~~\XTilde{Y}_l $ $\gets$ LS\_Y( $\XOmegaTilde{T},\ \ \XOmegaTilde{P}, \ \ \XTilde{X}_{l-1} , \ \ r $ ), \\ 
$~~~~~~~~~~~~~~\XTilde{X}_l $ $\gets$ LS\_X( $\XOmegaTilde{T}, \ \ \XOmegaTilde{P}, \ \ \XTilde{Y}_{l} , \ \ r $), \\
4:  \textbf{\ \ \ end for} \\
5:  $\ \ \tensor{X}$ $\gets$ ifft( $\XTilde{X}_L $, [], 3); $\tensor{Y} $ $\gets$ ifft( $\XTilde{Y}_L $, [], 3),  

\textbf{Output:} Pair of tensors ( $\tensor{X} $, $ \tensor{Y} $ ).    \\    
\rule{7.9cm}{0.12em}

sume that the full sampled seismic data volume has low-tubal-rank property and under-sampled
traces will increase the tubal rank. Therefore, the poststack 3D seismic data reconstruction can be tackled with tensor completion tools that using the low tubal rank property. The statement above transform to 
mathematical representation is that $\tensor{T}\in \Rset{R}{m}{n}{k}$ is a 3D seismic data with tubal-rank equal to $r$. Then, by
observing a set $\Omega \subset \Rset{R}{m}{n}{k}$ of $\tensor{T}$'s elements, we get the under-sampled seismic data $\tensor{T}_\Omega$. Then, our aim is to recover $\tensor{T}$ from  $\tensor{T}_\Omega$. The tensor reconstruction problem can be formulated as following

optimization function:
\begin{equation}
	\begin{aligned}
		\widehat{\tensor{T}} = \argmin{\tensor{X}}{m}{n}{k} \mathop{\mathrm{rank}}(\tensor{X}), \\
		\mathop{\mathrm{s.t.}} \ \ \POmega{\tensor{X}} = \POmega{\tensor{T}} .
	\end{aligned}
\end{equation}
Here, $\POmega{\cdot}$ denote the projection of a tensor onto the observed set $\Omega$, $\tensor{T}_\Omega = \POmega{\tensor{T}}$. The Tubal-Alt-Min 
algorithm proposed by Liu recently, which can complete the low tubal rank tensor with very high currency in several iterations 
is a perfect approach to solve this problem. 

\section*{Solution}
In the Tubal-Alt-Min algorithm, the target 3D seismic volume $\widehat{\tensor{T}} \in \Rset{R}{m}{n}{k}$ can be decomposed as 
$\widehat{\tensor{T}} = \tensor{X} \ast \tensor{Y}$, $\tensor{X} \in \Rset{R}{m}{r}{k}$, $\tensor{Y} \in \Rset{R}{r}{n}{k}$,
and $r$ is the target tubal-rank. With this decomposition, the problem (1) reduces to
\begin{equation}
	\widehat{\tensor{T}} = \argmin{\tensor{X}\in \Rset{R}{m}{r}{k}, \tensor{Y}}{r}{n}{k}\|\POmega{\tensor{T}} - \POmega{\tensor{X}\ast\tensor{Y}}\|_F^2.
\end{equation}

This cost function can be solved by the alternating minimization algorithm for low tubal rank tensor completion designed by Liu.
The main algorithm steps are showing as Alg 1. 

The key problem of Alg. 1 is the tensor least square minimization, which was solved by the providing methods in Liu's paper.
The main ideal is to decompose (2) into $n$ separate standard least squares minimization problem in the frequency domain. 
Then, we just need to solve a least square problem like the following form each step:
\begin{equation}
	\widehat{\vector{x}} = \argmina{\vector{x}}{rk}{1}~\|\vector{b} - \matrix{A_1}\matrix{A_2}\vector{x}\|_F^2.
\end{equation}
	
\section*{Performance evaluation}
To evaluate the performance of the algorithm we adopted, two commonly used evaluation criteria in seismic data completion filed  
have been used for comparison - the reconstruction error and the convergence speed.
\begin{itemize}
	\item Reconstruction error: here we adopted the relative squa-re error as a scale standard which defined as RSE $= \|\widehat{\tensor{T}} - \tensor{T}\|_F / \|\tensor{T}\|_F$.
\begin{comment}
the signal to noise ratio,
	\item Signal to noise ratio: to evaluate the denoising performance, we adopted the $SNR$, defined as
		 $SNR = 10*\log \frac{||\tensor{T}(:)||_2^2}{||\widehat{\tensor{T}}(:) - \tensor{T}(:)||_2^2}$.
		  Here, $\tensor{T}$ is noise-free data, $\widehat{\tensor{T}}$ is the denoised data.
	\item Running time: varying the tensor size and fixing other parameters, we measure cpu time in seconds.
\end{comment}
		\item Convergence speed: we measure decreasing rate of the RSE across iterations by linearly fitting the
			measured RSEs. 
\end{itemize}
In order to have an intuitive performance comparison, we compared with two other seismic data volume completion algorithm.
The Parallel matrix factorization algorithm (PMF)  \cite[]{Gao2015Parallel} and the  tensor
singular value decomposition based algorithm, also called tensor nuclear norm algorithm (TNN) \cite[]{Ely20155D} for seismic reconstruction. 
We applied those algorithm both on synthetic and real seismic data, the following subsections will 
demonstrate the detail performance comparison.

\subsection{Synthetic data example}
We use a Ricker wavelet with central frequency of 40 Hz to generate a simple 3D seismic model with two dipping planes. The seismic data 
corresponds to a spatial tensor of size $64 \times 64 \times 256$, 256 time samples with the time sampling rate of 1 ms and 64 corresponding to inline and xline direction. Then, through tSVD decomposition we get  $\tensor{U}$, $\tensor{S}$, and $\tensor{V}$. 
According to the decomposition result, we choose the first 2 tubes of $\tensor{S}$, and make other tubes elements equate to zeros, form a new 
tensor $\widehat{\tensor{S}}$. Then we get the tensor $\widehat{\tensor{T}} = \tensor{U} \ast \widehat{\tensor{S}} \ast \tensor{V}$ which tubal-rank is 2.

\begin{figure}[H]
	\vspace{-0.4cm}
	\centering
	\includegraphics[width=8cm]{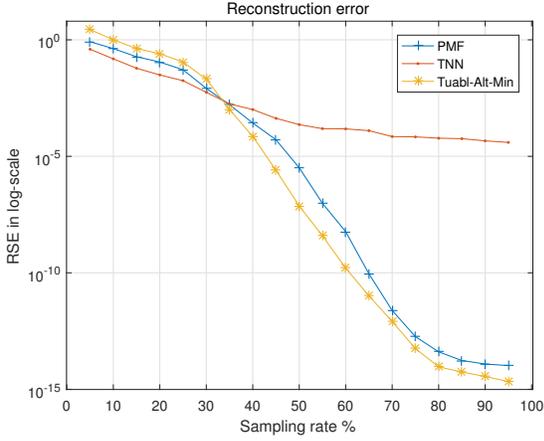}
	\caption{ Reconstruction error for the synthetic data set as a function of the downsampling rate.}
\end{figure}

\begin{figure}[H]
	\vspace{-0.4cm}
	\centering
	\includegraphics[width=8cm]{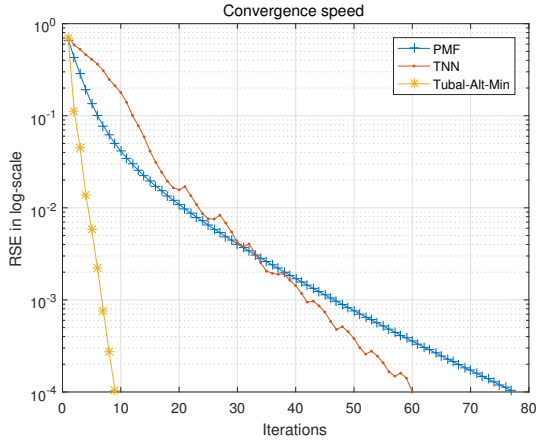}
	\caption{Convergence speed for the three algorithms on synthetic data.}
\end{figure}

Firstly, we apply Algorithm 1 to decompose the low-tuabl-rank tensor of our 
data set at different sampling rate vary from 5\% to 95\% and set the tubal-rank equate to 2. Using these
decompositions, we generate our reconstructed data and compare the error between the low-tubal-rank
reconstruction and full sampled data. Applying two other algorithm we stated above to reconstruct the data at 
the same condition. Because of all of the three algorithms include random sampling operator, we averaged those algorithms' 
performance under 20 above experiments. Figure 2 shows the relative square error (RSE) of the three algorithms. From these 
curves, we see that the performance of our algorithm is very outstanding. Even there are only 
30\% sampling traces, it can also get a relative perfect reconstruction data. As the increase of sampling rate, the reconstruction 
error of Tubal-Alt-Min algorithm decrease rapidly. Comparing with TNN, when the sampling rate over 50\%, it improves the recovery
 error by orders of magnitude. It's recovery error also better than the other algorithm almost in the whole sampling rate range.

Secondly, to evaluate the reconstruction error, we fixed the sa-
\begin{figure}[H]
	\addtocounter{figure}{0}
	\centering
	
	\setlength{\abovecaptionskip}{0.cm}
	\setlength{\belowcaptionskip}{0.cm}
	\vspace{-0.6cm}
	\subfigure[]{
		\hfill
		\includegraphics[height=5cm, width=3.5cm, viewport=1.0cm 1.0cm 12cm 15cm]{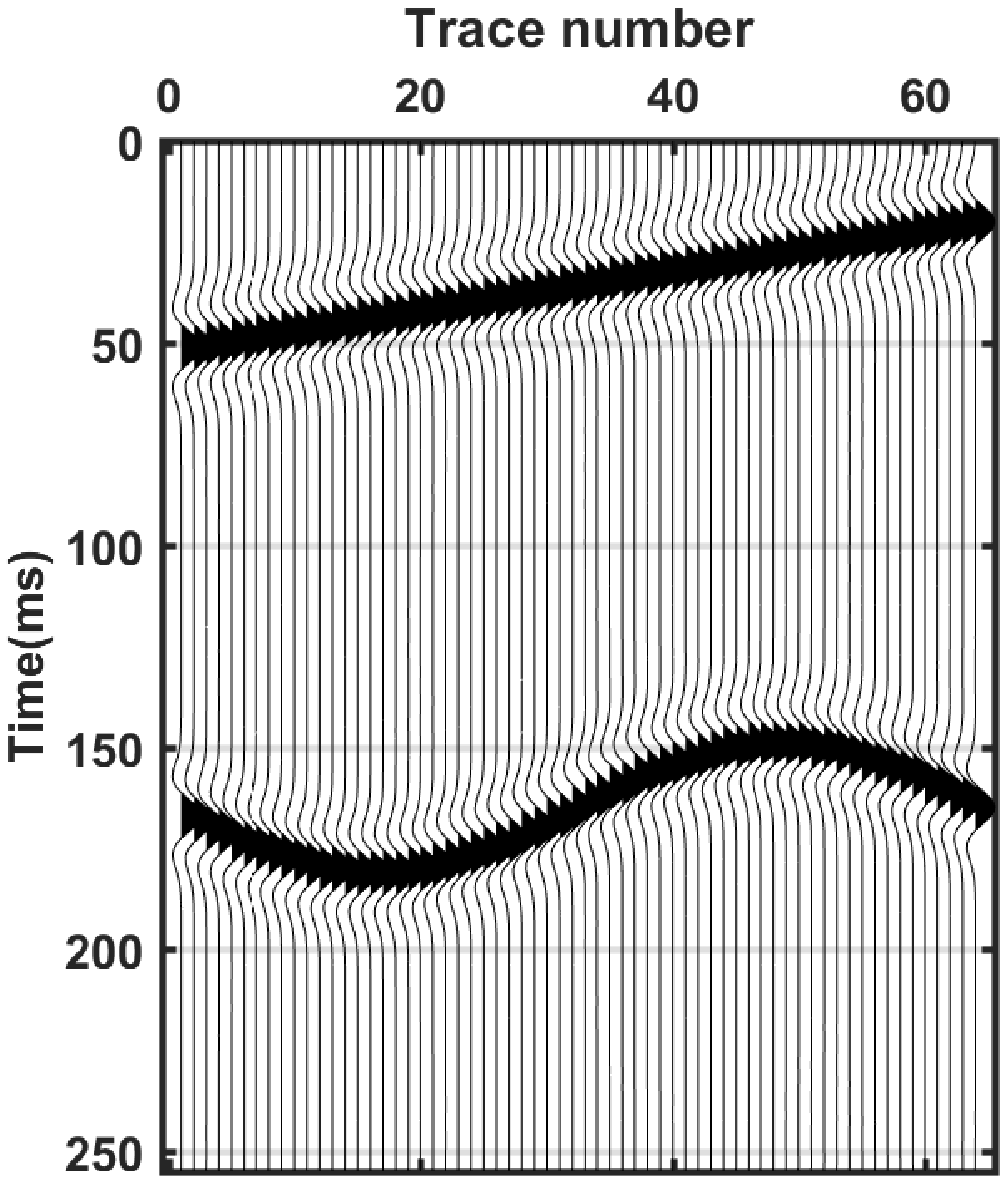}
	}
	\vspace{-0.6cm}
	\subfigure[]{
		\centering
		\includegraphics[height=5cm, width=3.5cm, viewport=1.0cm 1.0cm 12cm 15cm]{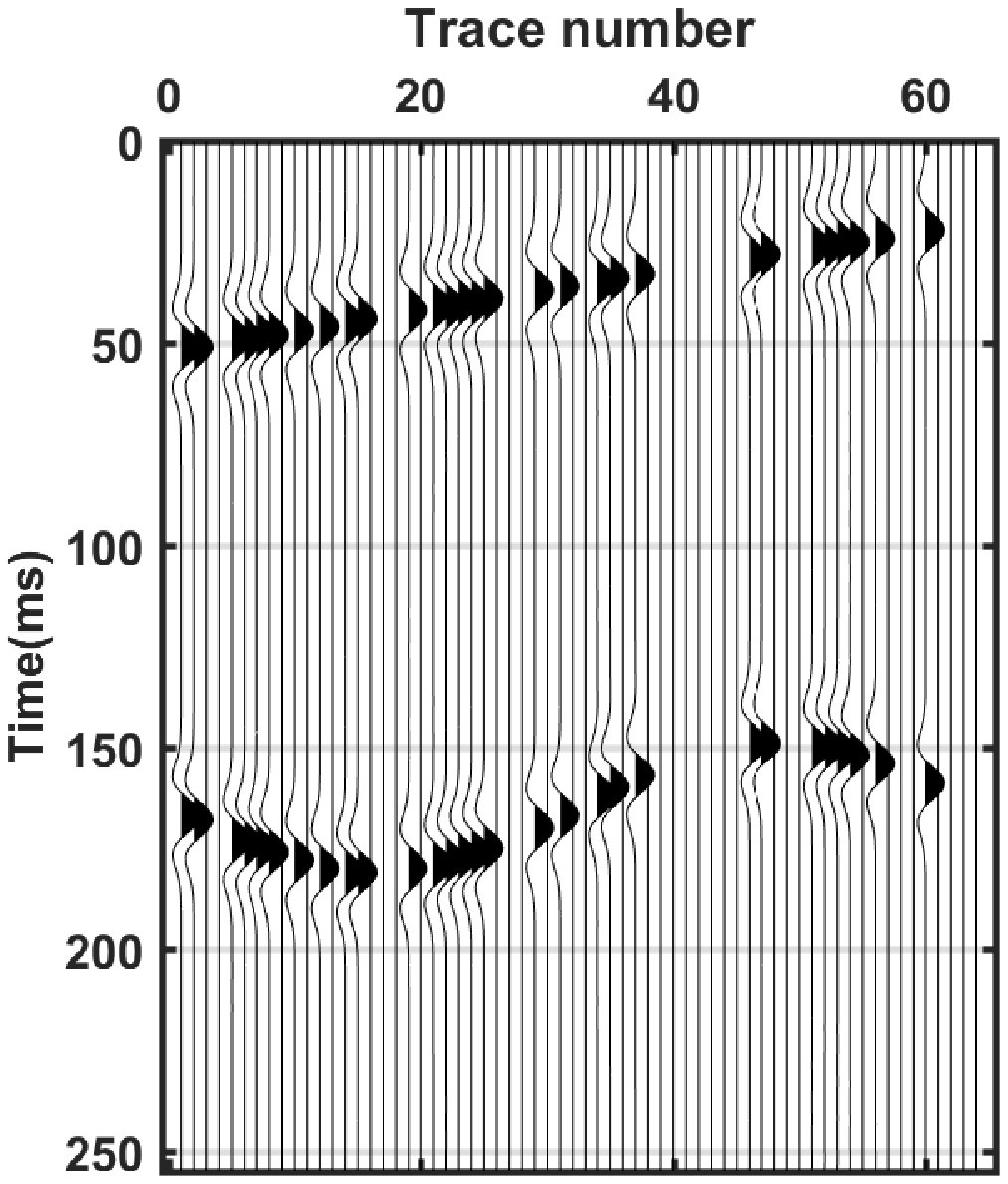}
	}
	\subfigure[]{
		\centering
		\includegraphics[height=5cm, width=3.5cm, viewport=1.0cm 1.0cm 12cm 15cm]{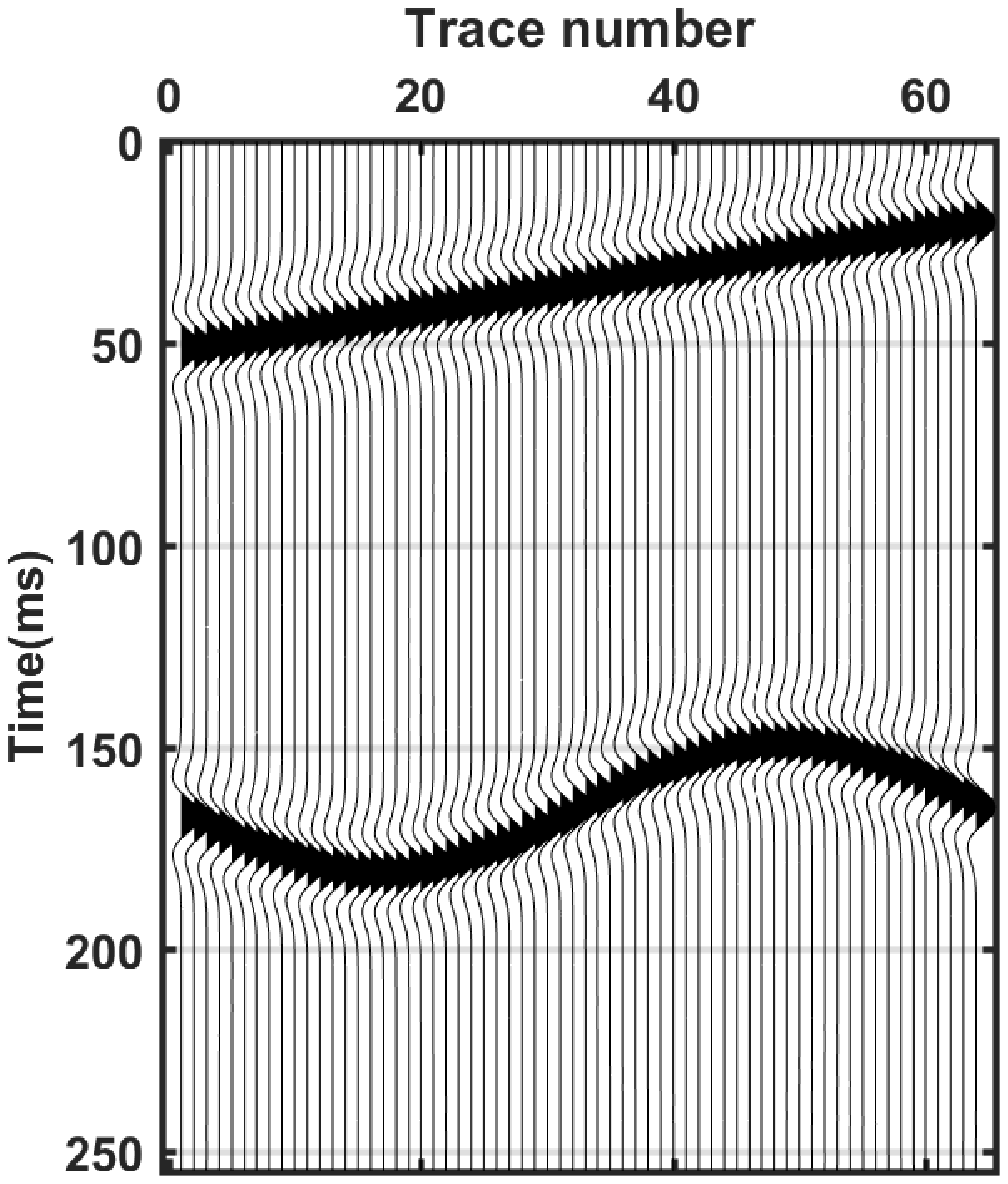}
	}
	\subfigure[]{
		\centering
		\includegraphics[height=5cm, width=3.5cm, viewport=1.0cm 1.0cm 12cm 15cm]{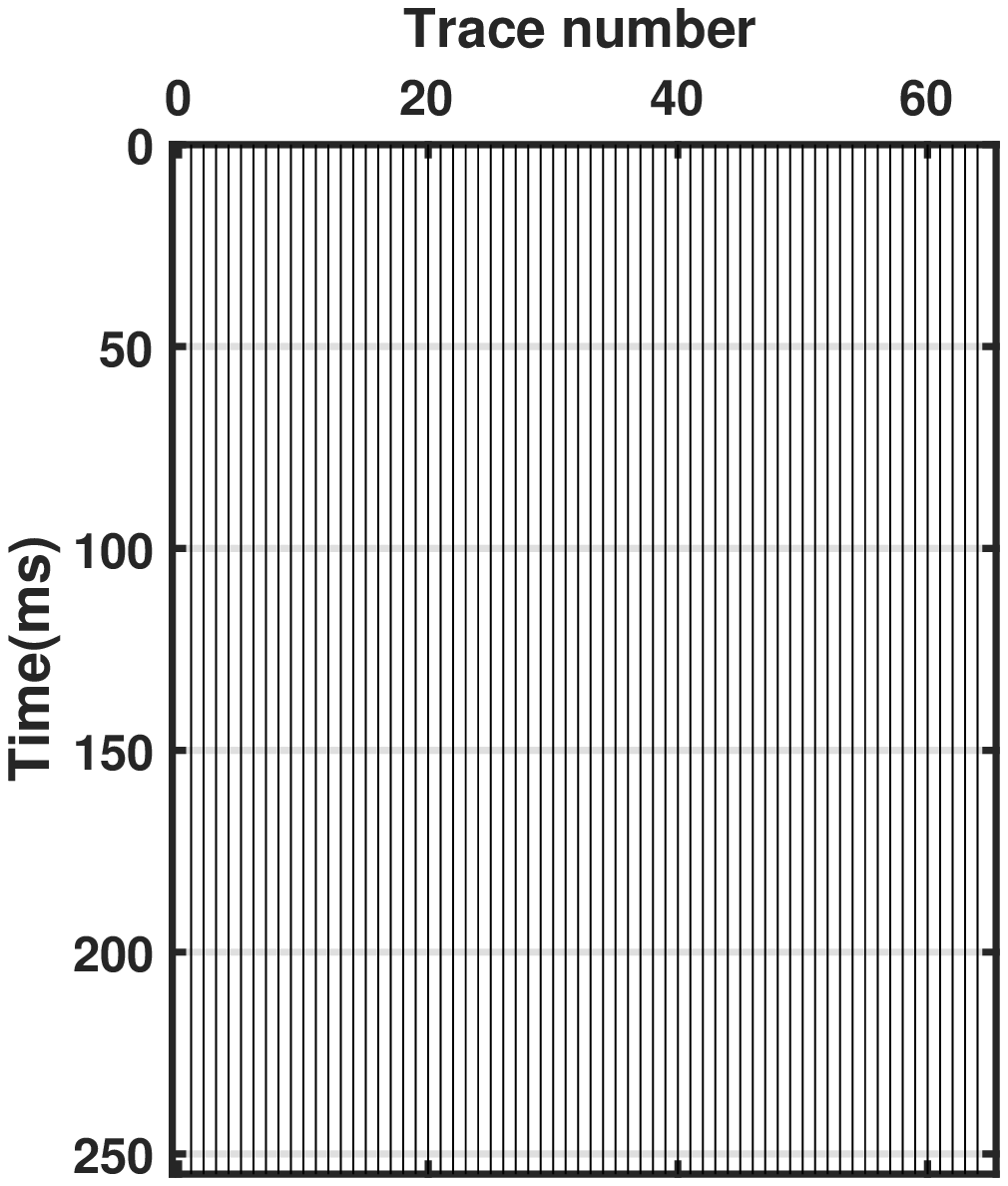}
	}
	\caption{(a) Full sampled synthetic data. (b) Under-sampled measured data for the case when 40\% of the traces in (a) was removed.  (c) The recovered traces of (b), it's RSE$ < 1e-4$. (d) The relative residual between (c) and (a), when sampling rate equal to 40\%, there almost no residual between original data and reconstruction data.}	
\end{figure}

mpling rate to be 40\%  and set the RSE tolerance of the three algorithms equal to 1e-4.
Then evaluate relative square error and convergence speed for the three algorithms. Figure 3 shows that Tubal-Alt-Min convergence at 9th iteration, 
TNN-ADMM terminates at the 60-th iteration, PMF used 77-th iterations to reach the preset threshold. The convergence speed of our algorithm is obviously better 
than other algorithms.

Then, we fixed the sampling rate at 40\% to evaluate the completion performance on seismic data. 
The portion of reconstruc-tion result shows in figure 4. From figure 4 (b), 4 (c) 
and 4 (d), we observe that the missing traces
 are effectively recovered.

\subsection{Filed data example}
We also test the performance of the reconstruction method on a
land data set that was acquired to monitor a heavy oil field in Anyue mountain, China.  
Our test data is a $25 \times 200 \times 600$ tensor cut from the full-sampled area of the filed data.
On this basis, we randomly sample 50\% traces in the seismic tensor, then reconstruct it. To reconstruct 
the traces, we evaluated the low-tubal-rank of the data manually. Determined by the tSVD, we found it's first 16 eigentubes's $l_2$ norm is 
much larger than the rest in almost 40 times. Consequently, we set the parameter of Alg. 1 $r = 16$. Figure 4 shows portion of the result cut 
along inline direction. From the comparison of figure 5 (a), 5 (c) and
\begin{figure}[H]
	\setcounter{subfigure}{0}
	\addtocounter{figure}{1}
	\centering
	\vspace{-0.4cm}
	\subfigure[]{
		\centering
		\includegraphics[height=6cm, width=3.5cm, viewport=1.0cm 1.0cm 11cm 14cm]{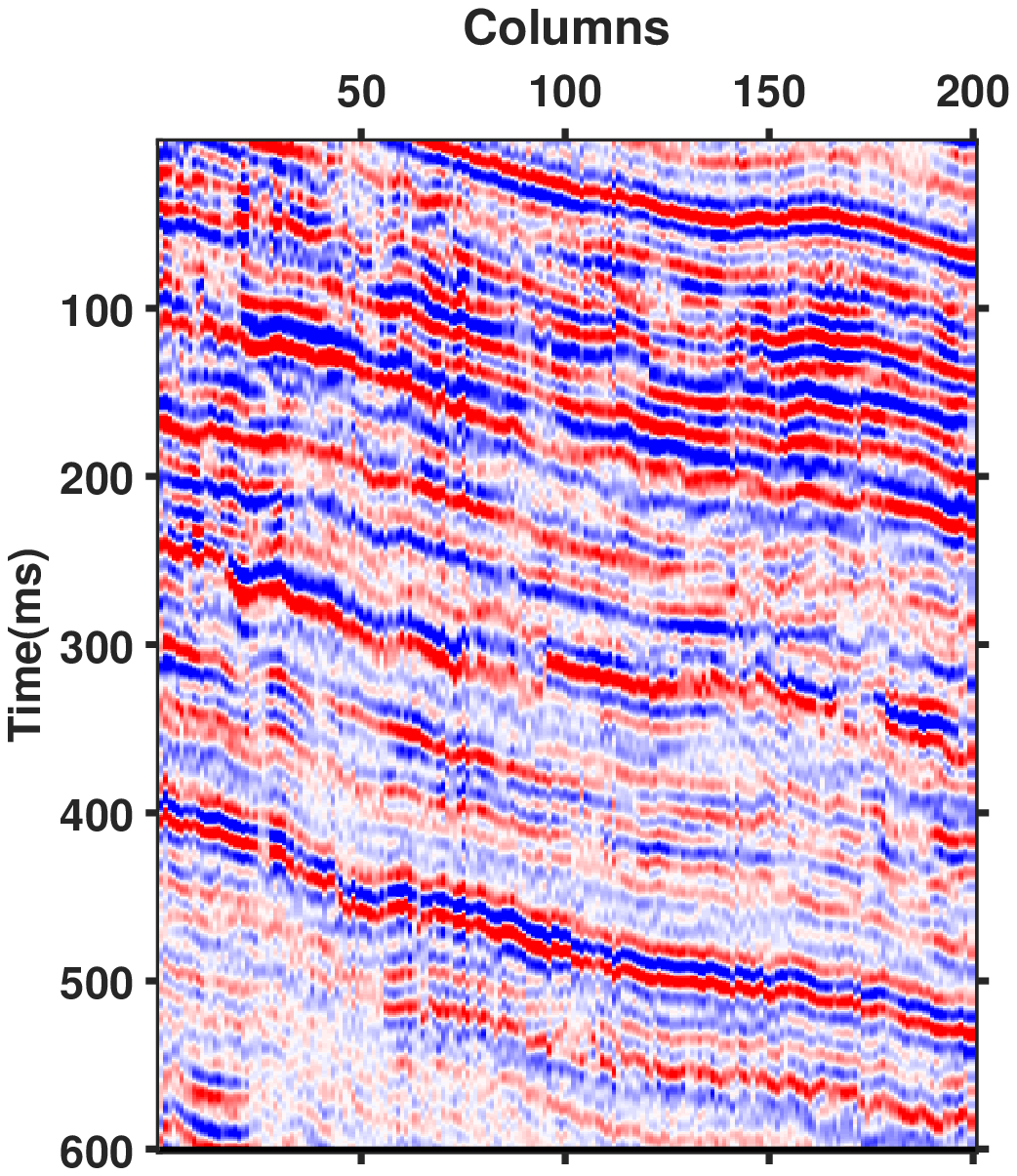}
	}
	\vspace{-0.4cm}
	\subfigure[]{
		\centering
		\includegraphics[height=6cm, width=3.5cm, viewport=1.0cm 1.0cm 11cm 14cm]{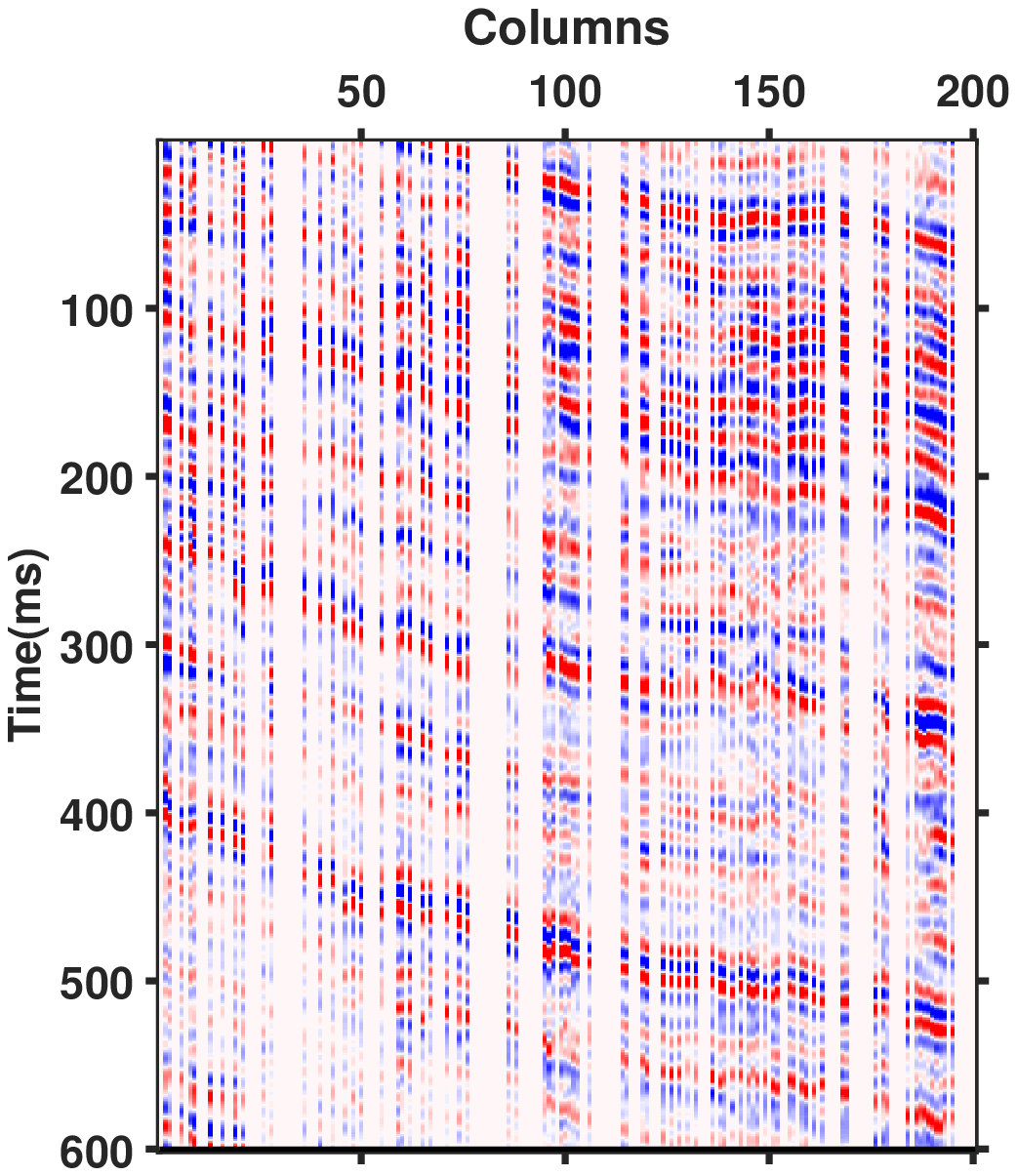}
	}	
	\vspace{0cm}
	\subfigure[]{
		\centering
		\includegraphics[height=6cm, width=3.5cm, viewport=1.0cm 1.0cm 11cm 14cm]{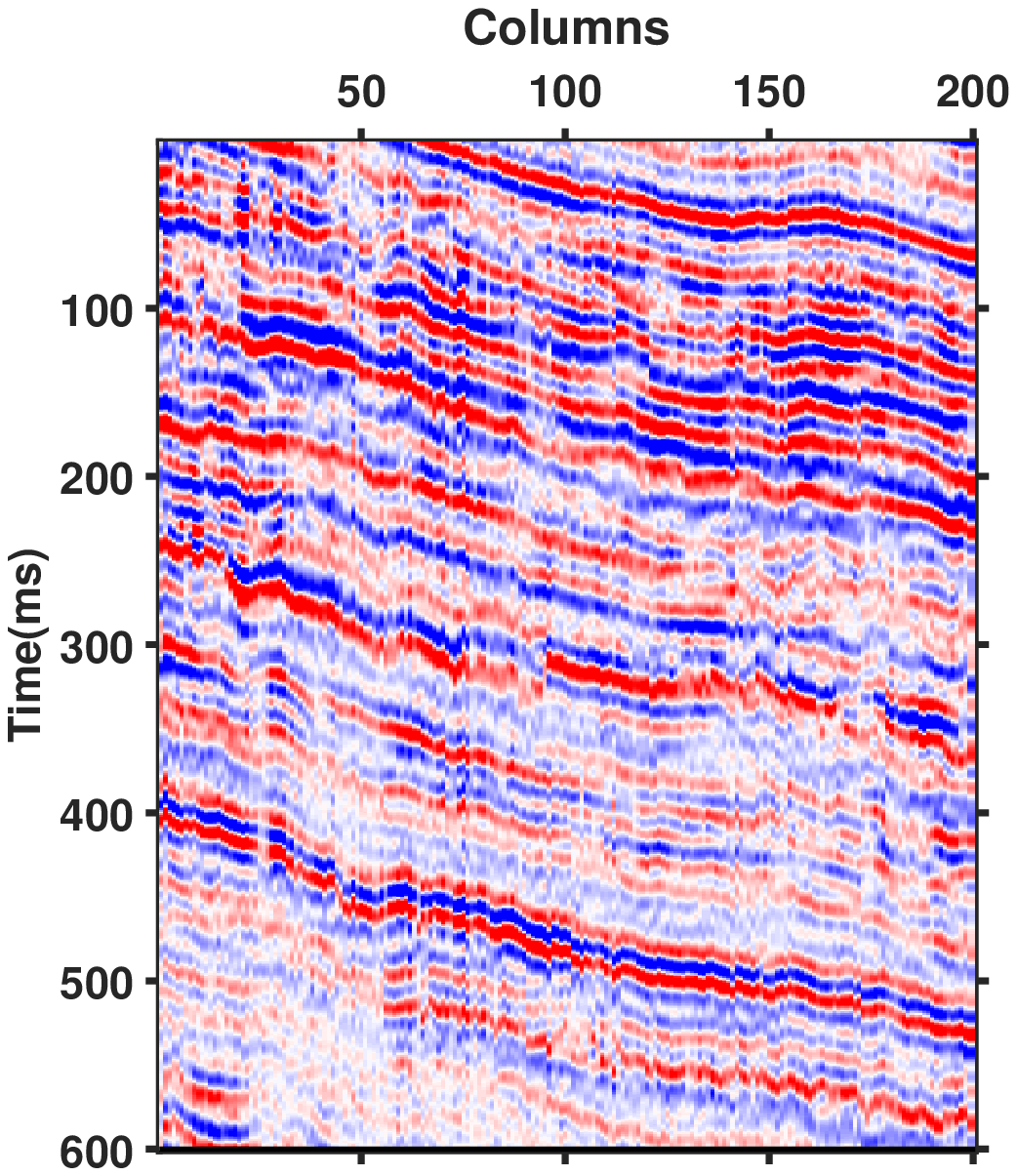}
	}	
	\vspace{0cm}
	\subfigure[]{
		\centering
		\includegraphics[height=6cm, width=3.5cm, viewport=1.0cm 1.0cm 11cm 14cm]{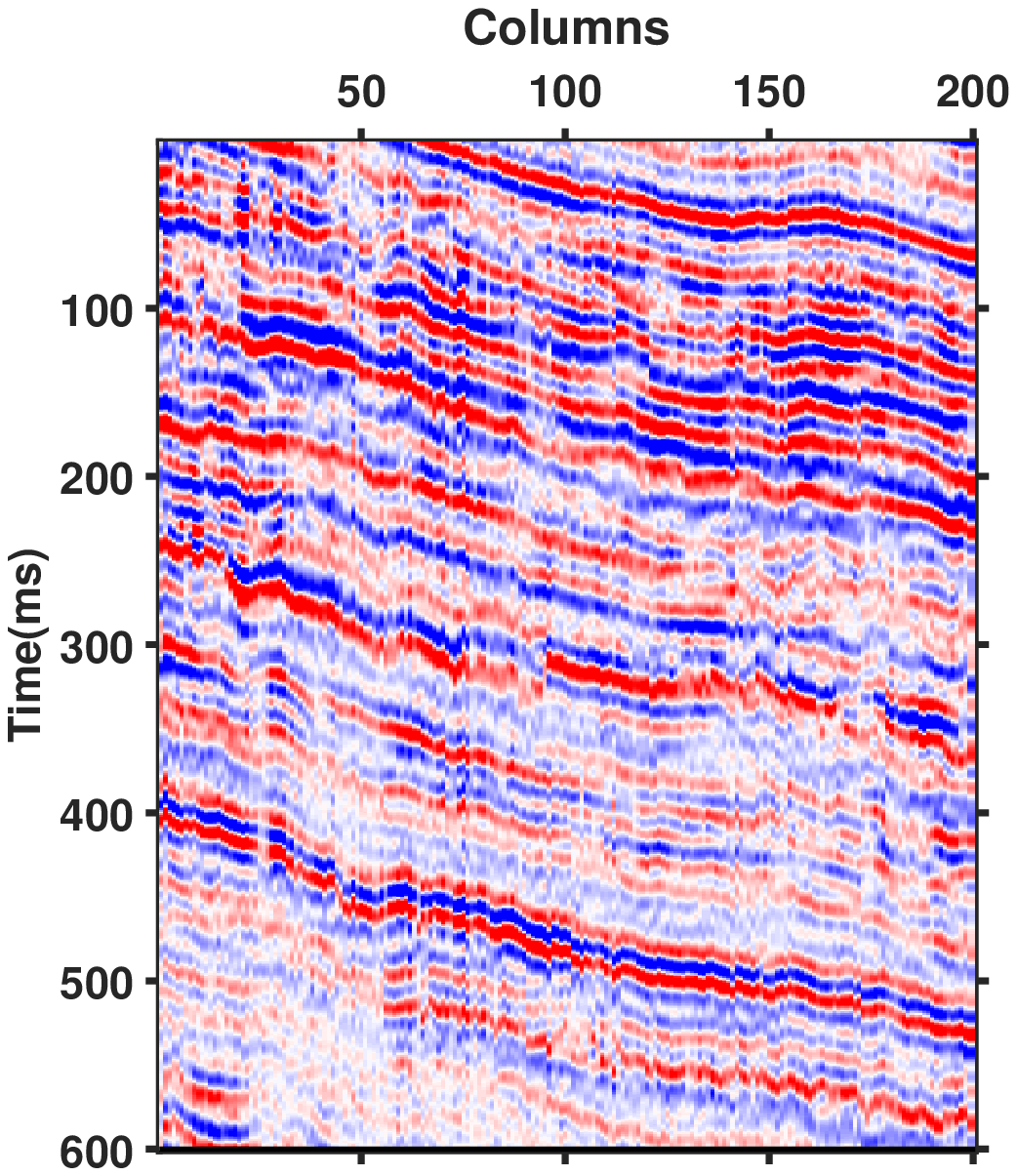}
	}	
	\caption{The reconstruction result of filed data. (a) The horizontal-slice of filed data from Anyue survey areas. (b) Randomly sampled 40\% trances from (a). 
		(c) The data recovered by Tubal-Alt-Min algorithm, it's RSE reached 1e-7. (d) The data recovered by TNN whose RSE converged to 1e-3.}
\end{figure}

 5 (d), the reconstruction performance is such amazing that almost no residual between 5 (a) and 5 (c)
Compared with the reconstruction result of TNN, it seems no difference. However, 
the RSE gap between them almost have a few orders of magnitude. Figure 5 shows the difference between the specific 
details of the seismic data traces restored by the two algorithms.

\begin{comment}
\begin{figure}[H]
	\addtocounter{subfigure}{0}
	\centering
	\ContinuedFloat
	\setlength{\abovecaptionskip}{-1.cm}
	\setlength{\belowcaptionskip}{-1.cm}

	\subfigure[]{
		\centering
		\includegraphics[height=5cm, width=3.5cm]{Fig/randomNoise.eps}
	}
	
	\subfigure[]{
		\centering
		\includegraphics[height=5cm, width=3.5cm]{Fig/Reconstruct.eps}
	}
	
\end{figure}

\begin{figure}[H]
	\setlength{\abovecaptionskip}{0.cm}
	\setlength{\belowcaptionskip}{-0.cm}
	\addtocounter{subfigure}{0}
	\centering
	\ContinuedFloat

	\subfigure[]{
		\centering
		\includegraphics[height=5cm, width=3.5cm]{Fig/RelateResidualnoise.eps}
	}
	\caption{(a) is the full sampled synthetic data. (b) randomly sample 40\% traces from (a). (c) randomly sample 40\% traces from (a), then add random noise make it's $SNR = 1$.   (d) is the recovered traces of (b), it's $RSE < 1e-7$. (e) is the recovered denoising traces of (c), it's $RSE < 1e-5$.
		(f) is the relative residual between (d) and (a). (g) is the relative residual between (e) and (a),  }	
\end{figure}
\end{comment}

\section*{Conclusion}
In this article, we formulate the problem of poststack seismic data reconstruction as 
an low-tubal-rank decomposition problem. The adopted method is based on the alternating minimization approach for low-tubal rank tensor completion. 
The unknown low-tubal-rank tensor is parameterized as the product of two much smaller tensors with the low-tubal-rank
 property being automatically incorporated, and Tubal-Alt-Min alternates between estimating those two tensors 
 using tensor least squares minimization. The biggest advantage of our algorithm is that it can achieve very high recovery accuracy in several it-
 
 \begin{figure}[H]
 	\setcounter{subfigure}{0}
 	\centering
 	\vspace{-0.7cm}
 	\subfigure[]{
 		\centering
 		\includegraphics[height=4.5cm, viewport=8cm 1.0cm 17cm 13cm]{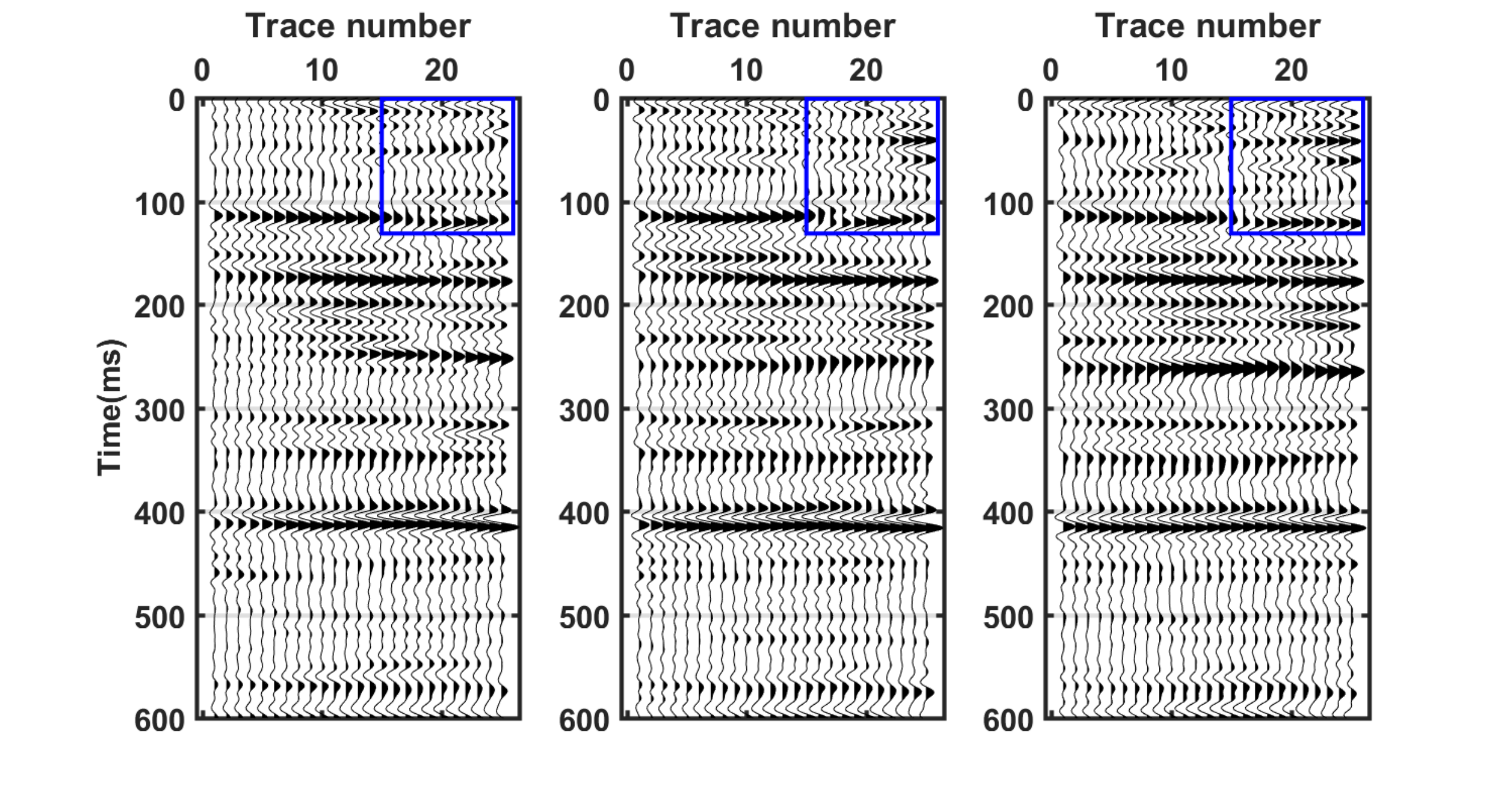}
 	}		
 \end{figure}
 
 \begin{figure}[H]
 	\centering
 	\vspace{-0.7cm}
 	\subfigure[]{
 		\centering
 		\includegraphics[height=4.5cm, viewport=8cm 1.0cm 17cm 13cm]{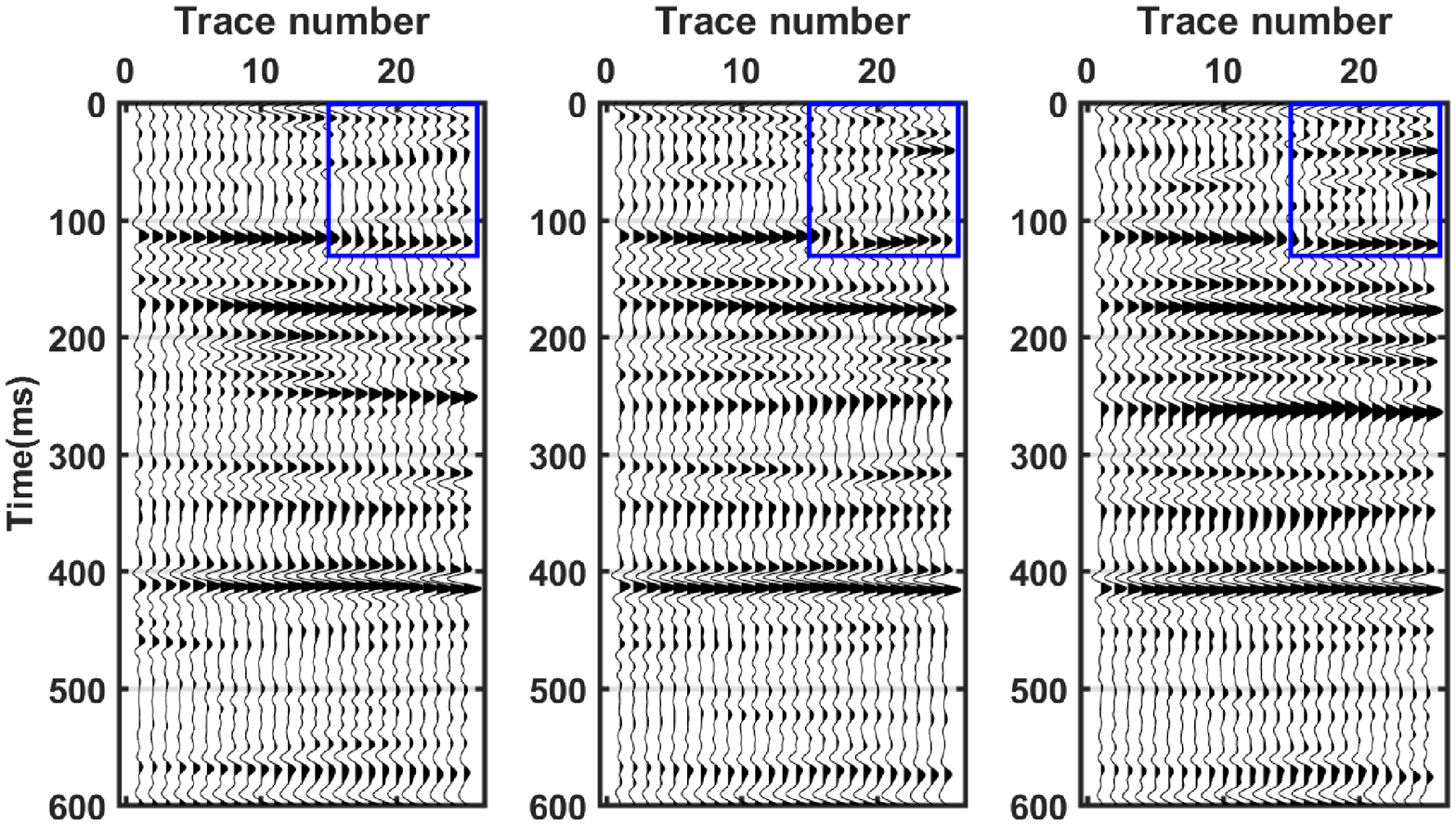}
 	}		
 \end{figure}
 \begin{figure}[H]
 	\addtocounter{figure}{0}
 	\centering
 	\vspace{-0.7cm}
 	\subfigure[]{
 		\centering
 		\includegraphics[height=4.5cm, viewport=8cm 1.0cm 17cm 13cm]{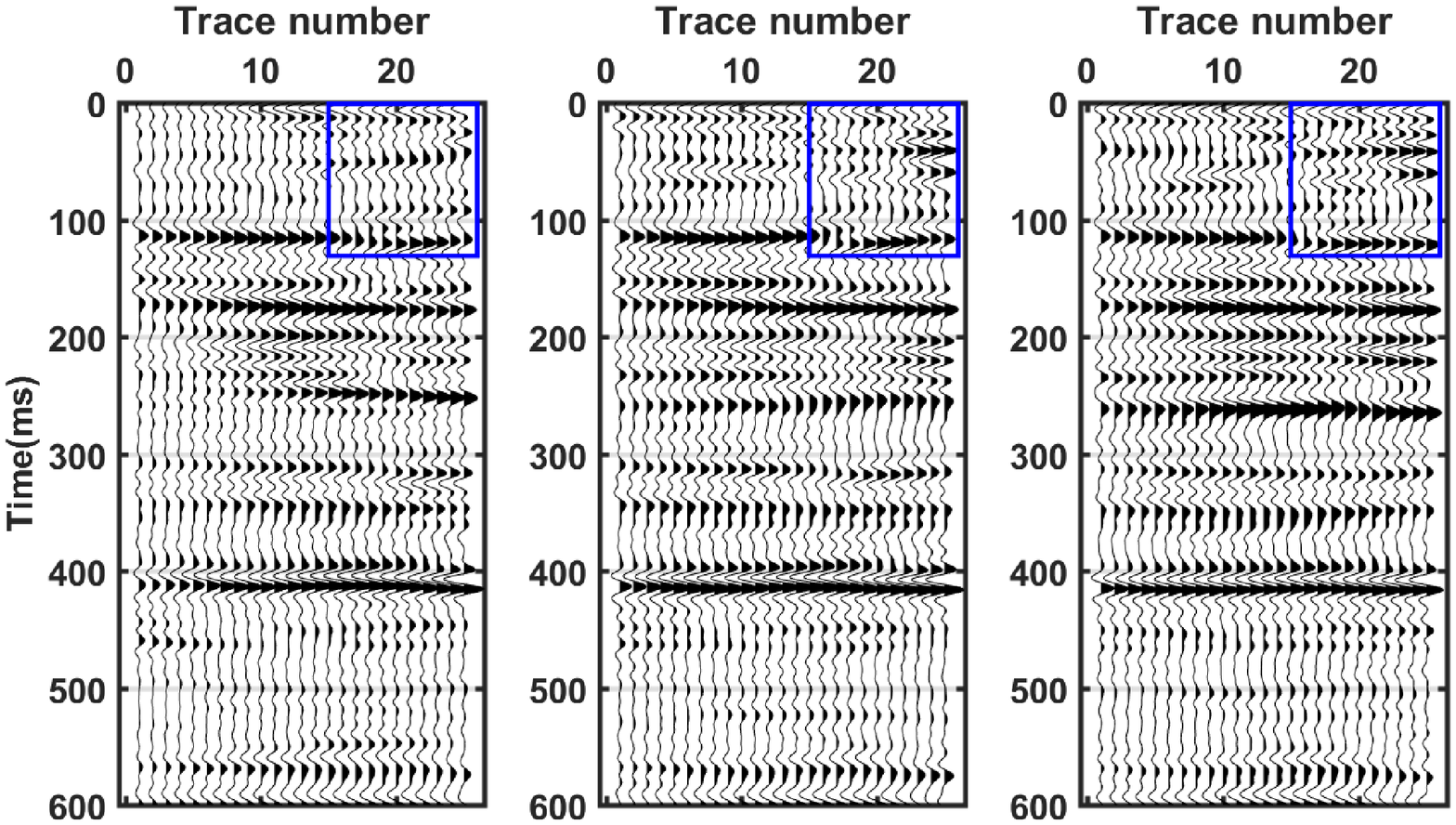}
 	}	
 	\caption{The reconstruction traces of filed data. (a) The lateral-slices of filed data from Anyue survey areas. (b) Traces recovered by TNN. 
 		(c) Traces recovered by our algorithm. From the recovered traces in the blue rectangles, our algorithm has an obvious performance advantages, 
 		which can recover more details of the seismic traces.}
 \end{figure}
 
 erations.   
 From the experimental result, it potently proved that compared with contrast algorithms our algorithm improves the recovery error by orders of magnitude 
 with much better convergence speed for higher sampling rates. 

\section{ACKNOWLEDGMENTS}
We would like to acknowledge financial support from the National
Natural Science Foundation of China (Grant No.U1562218).

\onecolumn

\bibliographystyle{seg}  % style file is seg.bst
%\bibliography{example}
% ref
\newcommand{\SortNoop}[1]{}

\end{document}